\documentclass[12pt,twoside]{article}   
\usepackage[super,sort,comma]{natbib}

\usepackage{fancyhdr}

\usepackage[section]{placeins}   %

\usepackage{graphicx}
\usepackage{dsfont}

\makeatletter \renewcommand\@biblabel[1]{$^{#1}$} \makeatother
\newcommand{\note}[1]{\mbox{}\\ \noindent \rule{16cm}{0.5mm} \\
{\em #1} \\ \noindent \rule{16cm}{0.5mm}
\typeout{    }
\typeout{***********note active on this page *************************}
\typeout{Note: #1  }
\typeout{****************************************end Note}
}

\newcommand{\cen}[1]{\begin{center} #1 \end{center}}

\usepackage[mathlines]{lineno}

\usepackage{hyperref}
\hypersetup{ colorlinks,
    citecolor=blue,
    filecolor=blue,
    linkcolor=blue,
    urlcolor=blue
}

\usepackage{xcolor}
       
\definecolor{gray}{rgb}{0.6,0.6,0.6}
\definecolor{red}{rgb}{0.85,0,0}
\definecolor{green}{rgb}{0,0.85,0}
\definecolor{blue}{rgb}{0,0,0.85}
\definecolor{beige}{rgb}{0.92,0.87,0.78}
\usepackage[all]{hypcap}    %causes link to figures to go to figure, not caption
\begin{document}

\cen{\sf {\Large {\bfseries Seven-probe scintillator dosimeter for treatment verification in HDR-brachytherapy} \\  
\vspace*{10mm}
Mathieu Gonod$^1$, Miguel Angel Suarez$^2$, Carlos Chacon Avila$^2$, Vage Karakhanyan$^2$, Clément Eustache$^2$, Samir Laskri$^3$, Julien Crouzilles$^3$, Jean-François Vinchant$^3$, Léone Aubignac$^1$ and Thierry Grosjean$^2$} \\
$^1$Centre Georges François Leclerc (CGFL) - Dijon, France\\
$^2$University of Franche-Comt\'e, CNRS, FEMTO-ST Institute, UMR 6174, Besançon, France\\
$^3$SEDI-ATI Fibres Optiques, 8 Rue Jean Mermoz, 91080 Évry-Courcouronnes, France\\

\vspace{5mm}
Version typeset \today}

\pagenumbering{roman}
\setcounter{page}{1}
\pagestyle{plain}
Author to whom correspondence should be addressed. email: thierry.grosjean@univ-fcomte.fr \\

\begin{abstract}

\noindent {\bf Background:} In vivo dosimetry (IVD) is gaining interest for treatment delivery verification in HDR-Brachytherapy. Time resolved methods, including source tracking, have the ability both to detect treatment errors in real time and to minimize experimental uncertainties. Multiprobe IVD architectures holds promise for simultaneous dose determinations at the targeted tumor and surrounding healthy tissues while enhancing measurement accuracy. However, most of the multiprobe dosimeters developed so far suffer either from compactness or inter-probe cross-talk issues.\\

\noindent {\bf Purpose:} We introduce a novel concept of a compact multiprobe scintillator detector and demonstrate its applicability in HDR-brachytherapy. Our fabricated seven-probe system is sufficiently narrow to be inserted in a brachytherapy needle or in a catheter.\\

\noindent {\bf Methods:} Our multiprobe detection system results from the parallel implementation of a miniaturized scintillator detector at the end of a bundle of seven fibers.  The resulting system, which is narrower than 320 microns, is tested with a MicroSelectron 9.1 Ci Ir-192 HDR afterloader, in a water phantom. The detection signals from all seven probes are simultaneously read with an sCMOS camera (at a rate of 0.06 s). The camera is coupled to a chromatic filter to cancel Cerenkov signal induced within the fibers upon exposure.  By implementing an aperiodic array of six scintillating cells along the bundle axis (one probe is kept bare to assess the stem effect), we first determine the range of inter-probe spacings leading to optimal source tracking accuracy. Then, three different source tracking algorithms involving sequentially or simultaneously all the scintillating probes are tested and compared. In each case, dwell positions are assessed from dose measurements and compared to the treatment plan. Dwell time is also determined and compared to the treatment plan. \\

\noindent {\bf Results:}  The optimum inter-probe spacing for an accurate source tracking ranges from 15 mm to 35 mm. The optimum detection algorithm consists of adding the readout signals from all detector probes. In that case, the error to the planned dwell positions is of 0.01 $\pm$ 0.14 mm and 0.02 $\pm$ 0.29 mm at spacings between the source and detector axes of 5.5 and 40 mm, respectively. Using this approach, the average deviations to the expected dwell time are of -0.006$\pm$0.009 s and  -0.008 $\pm$ 0.058 s, at spacings between source and probe axes of 5.5 mm and 20 mm, respectively.\\  

\noindent {\bf Conclusions:} Our seven-probe Gd$_2$O$_2$S:Tb dosimeter coupled to an sCMOS camera can perform time-resolved treatment verification in HDR Brachytherapy. This detection system of high spatial and temporal resolution provides a precise information on the treatment delivery via a dwell time and position verification of unprecedented accuracy. 
\end{abstract}
\note{This is a sample note.}

%\newpage     %may or may not be needed

\tableofcontents

\newpage

\setlength{\baselineskip}{0.7cm}      %double spacing		

\pagenumbering{arabic}
\setcounter{page}{1}
\pagestyle{fancy}

\section{Introduction}

High dose rate brachytherapy (HDR-BT) is a standard modality in cancer treatment which offers advantages of highly localized dose distributions and minimum number of treatment fractions. \cite{kubo:mp98,crook:sro20,shah:brachy20,viswanathan:brachy12}. To ensure that the planned dose is properly delivered, time-resolved in vivo dosimetry (IVD) has been proposed for monitoring treatments and detecting errors \cite{dewerd:mp11,valentin:icrp05,fonseca:piro20,verhaegen:phiro20,tanderup:mp13}. Among time-resolved IVD approaches, optical fibers coupled to scintillators have shown performances in time-resolved verification of the dose rate \cite{lambert:pmb06,lambert:mp07,therriault:mp11,andersen:mp09,belley:brachy18,johansen:pm19,kertzscher:ro11,jorgensen:mp21}, as well as dwell position and dwell time monitoring of a stepping radioactive source  \cite{johansen:brachy18,linares:mp20,johansen:pm19,jorgensen:mp21_2}, which represent clinically relevant information \cite{fonseca:piro20}.

IVD in multiprobe architectures has recently attracted attention for its ability to increase the spatial extent of treatment monitoring to volumes including the targeted tumor and surrounding healthy tissues. By use of individual detectors in various parallel catheters, Wang et al. \cite{wang:rm14} and Guiral et al. \cite{guiral:mp16} performed extended source tracking along the source catheter. Cartwright et al \cite{cartwright:mp10} realized a source tracking with an array of 16 plastic scintillator dosimeters embedded in a 20 mm-diameter rectal applicator. However, the acquisition rate of the detector was limited to 1 s, short dwell times of the source could not be assessed. Moreover, the diameter of the resulting multiprobe dosimeter in the centimeter range limits its field of application in BT. Therriault-Proulx at al developed a three-probe plastic scintillator detector  sufficiently narrow to be inserted within a BT needle or catheter \cite{therriault:mp13}, thereby making multiprobe tracking applicable in a broader range of BT. Because they involve one-millimeter outer diameter fibers, the three scintillator probes were engineered on the same fiber to be insertable into a BT needle or catheter. Signal demultiplexing at the fiber output was realized by use of a spectral filtering process of the light outcoming from the fiber multiprobe.  With a detection time of 3 seconds, the first prototype found limits in time-resolved monitoring of a treatment delivery. After optimization \cite{linares:mp19}, such a system showed highly improved performances in dose rate monitoring as well as dwell position and dwell time verification \cite{linares:mp20}. However, since the luminescence spectra of the three plastic scintillators noticeably overlap, the detector suffers from a cross-talk between its three detection channels, which may limit its accuracy.   

In this paper, we use the miniaturized scintillator detector (MSD) approach \cite{suarez:ox19,gonod:pmb21,gonod:pmb22} to demonstrate a seven-channel multiprobe detector that is narrow enough to perform time-resolved treatment monitoring within a single BT needle or catheter. Our 320-micron outer diameter device consists of 6 scintillating probes and a bare test-fiber engineered at the end of a narrow seven-fiber bundle. The parallel measurement of the seven readout optical signals at the bundle output with an sCMOS camera avoids inter-probe cross-talk and ensures a real-time IVD in a simple architecture (0.06 s detection rate). Each MSD of the detectiion system ensures minimum volume averaging within the steep dose gradients of BT sources, leading to unmatched source tracking performances in space and time. %\textcolor{red}{Despite its water nonequivalence,  the multiprobe detector is expected/shown to not perturbate the dose distribution in tissue equivalent media.} ??    

\section{Material and Methods}

\subsection{Multiprobe system}

\begin{figure}[htbp]
\begin{center}
\includegraphics[width=0.95\linewidth]{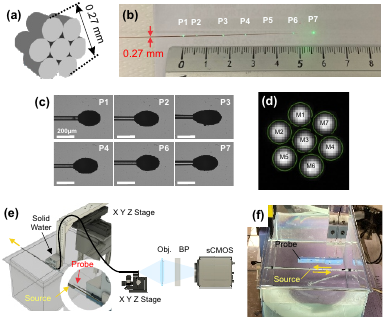}
\caption{(a) Scheme of the bundle of 7 fibers arranged in an hexagonal lattice. (b) Photograph of the multiprobe detector. Green laser light is coupled to the free bare facet of the fiber bundle to identify in the image the six scintillation cells (six green spots are observed due to light scattering of the fiber modes by the scintillators). (c) Magnified optical images of the six scintillation cells at the end of the fiber bundle. (d) Image of the bare face of the fiber bundle by the sCMOS camera when white light is projected onto the seven probes. (e) Schematics of the experimental setup involving a water tank, the multiprobe detector positioned onto a 2D motorized stage and a photometer based on an sCMOS camera coupled to an objective (Obj.) and a band-pass filter (BP).  (f) Photograph of the experimental setup.}\label{fig:exp}
\end{center}
\end{figure}

The multiprobe detector (MPD) shown in Fig. \ref{fig:exp}(b) involves a 10-meter-long bundle of seven biocompatible fibers  arranged in a hexagonal lattice (cf. Fig. \ref{fig:exp}(a); fabricated by SEDI-ATI). Each fiber is of 80-micron outer diameter (50-micron core diameter) and is covered with a 5-micron-thick polyimide protective coating. The total width of the bundle is of 270 microns. Each fiber tip is tapered in the form of a leaky-wave nano-optical antenna \cite{kraus:book,suarez:ox19} aimed at improving the transfer of the X-ray excited luminescence from the scintillators to the fiber. Scintillating powder (Gd$_2$O$_2$S:Tb) is selectively attached to the tapered tip of six of the seven fibers to form the probes P1-P4, P6 and P7 (see Fig. \ref{fig:exp}(b)). Gd$_2$O$_2$S:Tb is chosen as the scintillating material for its capability of a visible light emission with good efficiency, stability, linearity and temporal response \cite{qin:ox16,hu:saa18,alharbi:ieee18} and with very low sensitivity to temperature (in the range of 15$^{\circ}$-40$^{\circ}$) \cite{oreilly:ieee20}. The last bare fiber, labelled as P5, is used to evaluate the level of spurious Cerenkov signal generated within fibers upon irradiation. The overall fabrication process is detailed in Refs. \cite{gonod:pmb21,gonod:pmb22}. The scintillation cells forming the six parallel detectors are shown in Fig. \ref{fig:exp}(c).  The scintillation volume varies from 0.008 mm$^3$ (P1) to 0.009 mm$^3$ (P3). Four different inter-probe spacings along the bundle axis are defined by adapting the length of each optical fiber of the bundle. P1 and P2, P2 and P3, and P4 and P5 are spaced by $\Delta$, $2 \Delta$, and $4 \Delta$, respectively, where $\Delta=5$ mm, whereas P3 and P4 as well as P5 and P6 are both spaced by $1.7 \Delta$. The fiber bundle is positioned within a black 0.9-mm hytrel cladding to minimize collection of the background light from the test room. This opaque shield stops about 10 cm before the first probe P1 so that all six probes plus the bare test-fiber are directly in contact to water phantom. 

\subsection{Optical readout}

The optical signals at the end of the MPD are simultaneously recorded with an sCMOS camera (Andor Technology, Zyla model) whose maximum detection yield spectrally matches the emission of the Gd$_2$O$_2$S:Tb material. A 35 mm camera objective (Fujinon HF35SA) is positioned in front of the camera to image the bare output face of the fiber bundle at a rate of 0.06 s. Prior to acquisitions, we define seven regions of interest (ROI) tightly enclosing the seven light spots that are observable in the image (one spot per probe, see the green circles in Fig. \ref{fig:exp}(d) delimiting the ROIs). The image pixels located within each ROI are integrated to obtain seven detection signals sampled at 0.06 s.  A chromatic filter (544/24 nm band pass filter from Semrock) is positioned in front of the camera to filter out the spurious Cerenkov signal (stem effect) generated in the fibers upon exposure.  \cite{gonod:pmb21}. 

\subsection{Brachytherapy system}

A MicroSelectron afterloader with a 9.14 Ci Ir-192 HDR source (Air kerma strength of 37309 U) is used for irradiation. AAPM-TG43 protocol is used to calculate the reference dose rates applied to the treatment plan system \cite{perez:aapm12}. 

\subsection{Phantom}

The probe characterization is conducted in a 40x30x30 cm$^3$ water tank. The source catheter crosses the tank widthwise (Figs. \ref{fig:exp}(e) and (f)). During experiments, temperature in the water phantom varies from 17$^{\circ}$ to 19$^{\circ}$. The MPD is fixed to a solid-water holder that is attached to a 2D translation stage via a plastic adaptor (Figs.\ref{fig:exp}(e) and (f)). The fiber probe is set parallel to the source catheter. A coordinate frame of the set-up is defined so that the origin of the frame coincides with the scintillators of the proximal probe $P1$. The source-probe spacing along the (0x) and (0z) axes is determined with the motorized stage and the afterloader, respectively. 

\subsection{Detector specification and calibration}

The multiprobe detector being a parallel implementation of various MSD in a fiber bundle, the detector specification, in terms of linearity, repeatability and energy dependence can be found in Ref. \cite{gonod:pmb22}. 

The MPD is calibrated along seven lines parallel to the (0z)-axis. These lines are spaced by 5.5, 8, 10, 15, 20, 30 and 40 mm (along (0x)) from the axis of the source catheter. First, the MPD is positioned at the desired source-probe inter-catheter spacing $x$ with the motorized stage. Then, the source is displaced along the fixed source catheter by 2.5-mm steps. The calibration curves are obtained by integrating 165 images per source position. An interpolation (performed with Matlab software) is applied to the measured profiles to obtain a 0.1-mm sampling rate. During calibration, we verify with the scintillator-free fiber probe P5 that no optical signal (stem effect) is detected with the chromatic filter positioned in front of the camera.  

The signal-to-noise ratio (SNR) of the seven probes forming the detector is assessed at the dwell positions corresponding to the maxima of all the above-mentioned calibration curves, at the seven source-probe spacings $x$ ranging from 5.5 mm to 40 mm. The SNR is the average amplitude of the signal divided by its standard deviation.

\subsection{Dwell position and dwell time verification}

\subsubsection{Measurements}

To test the MPD, an irradiation protocole consisting of 40 dwell positions is applied for each value of source-probe spacing along $(0x)$. The dwell positions are spaced by 2.5 mm and the dwell time is fixed to 10 s. 

\subsubsection{Source position monitoring}

%Dwell position monitoring is realized at source-probe spacings $x$ below or equal to 40 mm. 
The instant position of the source at each acquisition time is retrospectively determined from the output signals of the MPD and the source activity, by use of the above-presented calibration curves. Since the displacement of the radioactive source between two successive dwell positions is operated over a few tens of milliseconds \cite{fonseca:mp15}, the resulting rise and fall times in the detected signal do not exceed two acquisition points. To ensure that these transitory phases are not taken into account in the source tracking, the first and the last signal points for each dwell position are ignored. 

During a treatment delivery, each probe $j$ ($1<j<7$) of the MPD delivers a temporal signal $S_j(t)$. At each instant $t=k \tau $, where $k \in \mathds{N}$ and $\tau$ is the acquisition time of the camera, the instant source position is deduced from the readout signal $S_j$ as follows.

First, function $f^k_j$ is defined for each probe $j$ as:

\begin{equation}
    f^k_j(z)=\left| C_j(z) - S_j^k \right|,
\end{equation}

where $C_j(z)$ and $S_j^k$ are the calibration curve and the readout signal of the $j^{th}$ probe at the $k^{th}$ time step, respectively. $z$ corresponds to the spatial coordinate along the axis of the source catheter. Calibration curves being symmetric regarding the $z$-coordinate, each probe provides two likely instant source positions located on both sides of function $f^k_j$. Therefore, at minimum two probes are necessary to unambiguously determine the position of a stepping BT-source. 

The instant source position $Z^k$ is determined using four different methods involving various manipulations of functions $f^k_j$. To find the inter-probe spacing which optimizes source tracking accuracy, source position verification is realized from a "two-probe" dosimetry using Eq. \ref{eq:z3} ($m=1$). $j_1 \in [1,6] \backslash {5}$ and $j_2 \in [2,7] \backslash {5}$ are the indices of the probes forming the 15 probe pairs ($j_1$,$j_2$) allowed by our detector. The inter-probe spacing ranges from 5 mm to 52 mm. Probe P5, which is bare to assess in-fiber Cerenkov effect, is not involved in the source position monitoring. Source tracking is systematically analyzed from each of the 15 probe pairs of the detector.  

\begin{equation}
    Z_m^k(z)= \min\left[f^k_{j_1}(z)+f^k_{j_2}(z)\right], \label{eq:z3}
\end{equation}

Three source tracking algorithms have also been tested and compared. In each case, all the six scintillating probes are involved in the source position monitoring during the treatment delivery. The two first values of the instant source position ($Z_m^k$, $m=$2 and 3) are calculated from the readout signals of probe pairs dynamically chosen among the seven available probes of the MSD. In both of these two-probe measurements, the source position is defined from Eq. \ref{eq:z3} with indices $j_1$ and $j_2$ which vary with the source position along (0z). Probe pairs are dynamically chosen to provide the higher readout signals ($m$=2) or on the basis of a maximum gradient of their calibration curves at the source position ($m$=3). %Source tracking from a single probe is supposed to be of maximum accuracy at the locations where the spatial gradient of probe's calibration curve is maximum. 
The last $z$-coordinate of the source $Z_4^k$ is determined by adding functions $f^k_j$ of all probes. We have:

\begin{equation}
    Z_4^k(z)= \min\left[\sum_{i=1}^7 f^k_j(z)\right], \label{eq:z1}
\end{equation}

\subsubsection{Dwell time verification}

The monitoring of an HDR-BT treatment is known to produce a staircase temporal signal \cite{belley:brachy18,johansen:pm19,gonod:pmb21}. The dwell times of the stepping source, which correspond to the duration of the plateaus in between two successive signal edges, can be simply determined from an edge detection within all readout signals of our MPD system.  Our edge detection approach involves function $F^k$ defined as: 

\begin{equation}
    F^k=\sum_{j=1}^6 \left(S_j^{k+1}-S_j^k \right), 
\end{equation}

By adding the signal derivatives from all probes, our algorithm is expected to reduce undesired fluctuations in the edge detection function (due to readout noise), as compared to that of a single probe detector.  

%\subsection{Dose rate retrieval}

%The dose rate ($D_{meas}$) is determined from the measured dwell position and source activity by applying the TG-43 formalism \cite{perez:aapm12}. The relative offset to the planned dose rate  $\Delta D / D = D_{meas}/D_{TPS}-1$ is then calculated. $D_{TPS}$ is the planned dose rate for each dwell position of the TPS.

\section{Results}

\subsection{Detector characterization and calibration}

\begin{figure}
\begin{center}
\includegraphics[width=0.6\textwidth]{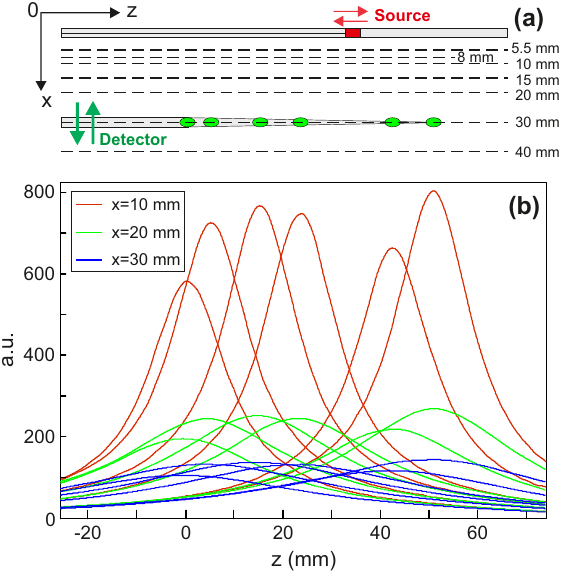}
\caption{(a) Schematic of the experimental set-up in the water tank. The coordinate frame defining the axis convention is shown in the top-left corner. The source and MPD are represented in red and green, respectively. All seven positions of the MPD defining source-probe spacings $x$ ranging from 5.5 to 40 mm are represented with dashed lines. The source and MPD move along the (0z) and (0x) axes, respectively. (b) Calibration curves of the MPD used for source tracking at $x$ equal to 10, 20 and 30 mm.}\label{fig:calib}
\end{center}
\end{figure}

An MPD being the parallel implementation of various MSD in a fiber bundle, the detector specification, in terms of linearity, repeatability and energy dependence can be found in Ref. \cite{gonod:pmb22}. Figure \ref{fig:calib}(b) shows the calibration curves of the MPD acquired at source-probe spacings $x$ of 10, 20 and 30 mm. Six gaussian-like profiles are shown per source-probe inter-catheter spacing (one profile per probe), whose maxima coincide with the probe positions along (Oz). The experimental configuration is schemed in Fig. \ref{fig:calib}(a). The SNR of the probes forming the detector varies from 110-140 down to 20-25 at source-probe spacings $x$ of 5.5 mm and 40 mm, respectively. 

\subsection{Source position monitoring}

\subsubsection{Optimal probe-to-probe spacing}

\begin{figure}
\begin{center}
\includegraphics[width=0.6\textwidth]{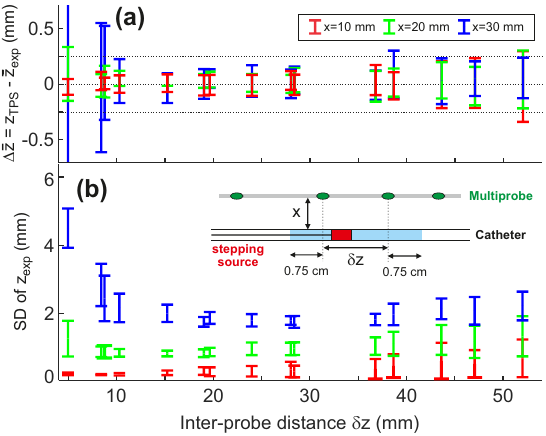}
\caption{Optimal inter-probe spacing for source tracking. (a) Offset between the measured and planned dwell positions along (0z) ($\bar{z}_{exp}$ and $z_{TPS}$, respectively) as a function of the inter-probe distance $\delta z$. The measured dwell position $\bar{z}_{exp}$ corresponds to the average of the instant source position $Z^k_1$ (see Eq. \ref{eq:z3}) over a dwell time. Each error bar shows for one source-probe inter-catheter spacing $x$ (see inset of (a)) the offset analysis to the planned dwell positions (mean and standard deviation). This analysis is performed over $z$-coordinates spanning over $\delta z \pm 0.75$ cm (see inset of (b)). (b) SD of the experimentally determined source position $Z^k_1$ versus the inter-probe distance $\delta z$. Each error bar shows the distribution (mean and standard deviation) of the SD for dwell positions within $\delta z \pm 0.75$ cm (see inset of (b)), at a given source-probe inter-catheter spacing $x$ (see inset of (a)).}
\label{fig:dpi}
\end{center}
\end{figure}

The aperiodic scintillator array of our MPD enables 15 probe pairs whose inter-probe spacings vary from 5 mm to 52 mm. To identify optimal inter-probe spacings for the future MPD designs, we analyzed the accuracy of the source position verification over these 15 probe pairs, versus the inter-probe distance $\delta z$ along the detector axis (0z) (see Fig. \ref{fig:dpi}).  The instant source position is here determined from $Z^k_1$ function, cf. Eq. \ref{eq:z3} ($m=1$). The displacement range of the BT source along (0z) varies with the inter-probe distance $\delta z$ as $\delta z \pm 0.75$ cm (see inset of Fig. \ref{fig:dpi}(b)).  

We see from Fig. \ref{fig:dpi}(a) that the minimum deviation to the planned dwell positions occurs at $\delta z$ values in-between 15 mm and 35 mm, regardless of the source-probe inter-catheter spacing $x$. In that $\delta z$ range, the offset distribution to the planned dwell positions does not exceed 0.05 $\pm$ 0.15 mm at $x$=30 mm ($0.15 \pm 0.41$ mm at $x$=40 mm). The SD of the measured  instant source position $Z^k_1$ reaches minimum values at $\delta z$ in between $0.87 x$ and $x$ (see Fig. \ref{fig:dpi}(b)). This property, which is observed for all values of $x$ ranging from 5.5 mm to 40 mm, is imputed to the broadening  of the calibration curves along (0z) as the source-probe inter-catheter spacing $x$ increases (cf. Fig \ref{fig:calib}). The tighter distributions of SD are of 0.2 $\pm$ 0.022 mm, 0.82$\pm$ 0.12 mm and 1.78 $\pm$ 0.12 mm at $x=$10, 20 and 30 mm, respectively. 

%Figure \ref{fig:dp}(a) shows the mismatch between the measured and planned dwell positions, as a function of the spacing $\delta z$ between the two selected probes. The measured dwell position $\bar{z}_{exp}$ corresponds to the average of the instant source position $z$ measured over a dwell time. 

%%%%%%%%%%%%%%%%%%%%%%%%%%%%%%%%%%%%%

\subsubsection{Optimal algorithm for source tracking: two-probe versus all-probe detection strategies}

\begin{figure}
\begin{center}
\includegraphics[width=0.6\textwidth]{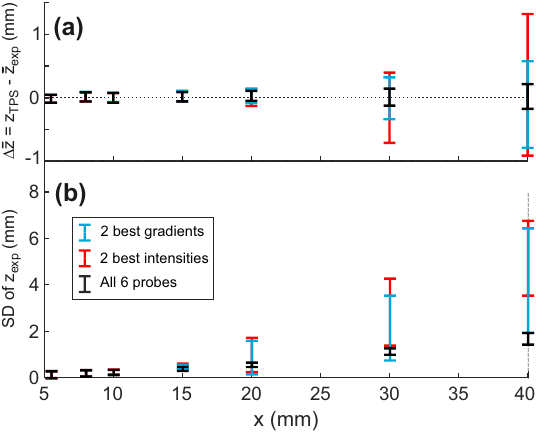}
\caption{Optimal algorithm for source tracking: two-probe versus all-probe detection approaches. (a) Offset between the measured and planned dwell positions along (0z) ($\bar{z}_{exp}$ and $z_{TPS}$, respectively) as a function of the source-probe inter-catheter spacing $x$. The measured dwell position $\bar{z}_{exp}$ corresponds to the average of the instant source position over a dwell time. The instant source position corresponds to $Z^k_2$, $Z^k_3$ or $Z^k_4$ (cf. Eqs. \ref{eq:z1} and \ref{eq:z3}; see inset of (b)). Each error bar shows for one measurement method the analysis of the mismatch to the planned dwell positions (mean and SD). This analysis is performed over a range of z-coordinates of 10 cm enclosing the detector. (b) SD of the experimentally determined instant source positions $Z^k_2$, $Z^k_3$ and $Z^k_4$. Each error bar shows for one measurement method (see inset of (b)) the analysis (mean and SD) of the distribution of SD for dwell positions spanning over 10 cm.}\label{fig:dp2all}
\end{center}
\end{figure}

A seven-probe detector enables numerous detection strategies for source tracking. In Fig. \ref{fig:dp2all}, we compare three algorithms which involve dosimetry either from probe pairs dynamically chosen during the treatment delivery (cf. $Z^k_2$ and $Z^k_3$ of Eq. \ref{eq:z3}) or from all the probes of the detector (cf. $Z^k_4$ of Eq \ref{eq:z1}). 

At source-probe spacings $x$ below 15 mm, all three methods provide the same source tracking accuracy. The mean and standard deviation of the offset to the planned dwell positions do not vary by more than 0.008 and 0.014 mm, respectively, from one method to another (Fig. \ref{fig:dp2all}(a)). With the two higher signal method (cf. $Z^k_2$ of Eq. \ref{eq:z3}), the deviation to the planned dwell positions, which is of 0.007$\pm$0.138 mm at $x$=20 mm, increases up to  0.20 $\pm$ 1.12 mm at $x$= 40 mm. As a comparison, the dwell position verification from the two signals of steeper gradients (calculation of $Z^k_3$) leads to a mismatch to the treatment plan of 0.027$\pm$0.115 mm and -0.11$\pm$0.68 mm at $x$ equal to 20 and 40 mm, respectively. Source tracking from all detected signals (calculation of $Z^k_4$) is much less impacted by the enhancement of the source-probe inter-catheter spacing $x$. The offset to the planned dwell positions is of 0.029$\pm$0.078 mm at $x$=20 mm and 0.02$\pm$0.19 mm at $x$=40 mm, respectively.

The SD of the instant source position determined from the three above-mentioned methods is reported in Fig. \ref{fig:dp2all}(b). For source-probe distances below 15 mm, all three methods determine the instant source positions with almost the same accuracy. When $x$ exceeds 20 mm, the dwell position verification from the two higher detected intensities (calculation of $Z^k_2$) is the less accurate. A detection from all probes (cf. $Z^k_4$) minimizes signal fluctuations. 

%%%%%%%%%%%%%%%%%%%%%%%%%%%%%%%%%%%%%

%Fig. \ref{fig:dp} displays a detailed representation of the source position determined from all six probes, which correspond to the analysis shown in Fig. \ref{fig:dp2all} (cf. black error bars). Fig. \ref{fig:dp}(a) reports the deviation to the planned dwell positions. Each point corresponds to the difference between the planned and measured dwell positions ($z_{TPS}$ and $\bar{z}_{exp}$, respectively). Fig. \ref{fig:dp}(b) reports the SD of the instant source position $Z^k_1$??, i.e., the SD of the distribution of source positions measured at a rate of 0.06s during a dwell time. Figure \ref{fig:dp}(a) shows that the offset to the planned dwell position remains at the same level over the entire displacement range of the stepping source (i.e., 10 cm), regardless of the source-probe spacing $x$.  On the contrary, noticeable enhancement of the SD is observed when the source is positioned out of the region where the probes are located, i.e., in between the detector and probe $P_1$ ($\bar{z}_{exp}$<0) or beyond probe $P_7$ ($\bar{z}_{exp}>$5.7 cm ; cf. Figs. \ref{fig:exp}(b) and \ref{fig:calib}). Such a local enhancement of the SD is maximum when $x$=5.5 mm and lessens as the source-probe spacing $x$ increases, to finally vanish at $x$=20 mm. 

\begin{figure}
\begin{center}
\includegraphics[width=1\textwidth]{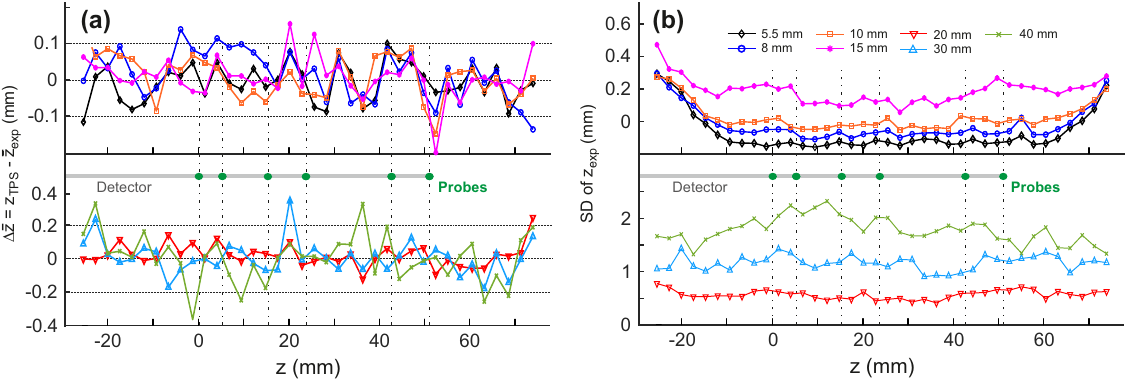}
\caption{Analysis of an "all-probe" source tracking. (a) Offset between the planned and experimentally determined dwell positions along (0z) ($z_{TPS}$ and $\bar{z}_{exp}$, respectively). $\bar{z}_{exp}$ is the average of the measured instant source position $Z^k_4$ over a dwell time. The mismatch to the planned dwell positions is shown for seven values of the source-probe inter-catheter spacing $x$ (see legend of (b)). (b) SD of the measured instant source position $Z^k_4$ along (0z) as a function of the z-coordinate. Here again, seven source-probe spacings $x$ are considered (see legend). Insets of (a) and (b), schematic of the multiprobe detector which identifies the $z$-coordinates of the scintillators on the graphs (with dashed lines).}\label{fig:dpa}
\end{center}
\end{figure}

Fig. \ref{fig:dpa} displays a detailed representation of the source position determined from all scintillating probes, which correspond to the analysis shown in Fig. \ref{fig:dp2all} (cf. black error bars). Fig. \ref{fig:dpa}(a) reports the error to the planned dwell positions, which corresponds to the difference between the planned and measured dwell positions ($z_{TPS}$ and $\bar{z}_{exp}$, respectively). Fig. \ref{fig:dpa}(b) reports the SD of the instant source position $Z^k_4$, i.e., the SD of the distribution of source positions measured at a rate of 0.06s during a dwell time. Noticeable enhancement of the SD is observed when the source is positioned out of the region where the probes are located, i.e., in between the detector and probe $P_1$ ($\bar{z}_{exp}<0$) or beyond probe $P_7$ ($\bar{z}_{exp}>5.2$ cm ; cf. Fig. \ref{fig:exp}(b) and Fig. \ref{fig:calib}). This local  fluctuation enhancement is maximum when $x$=5.5 mm and lessens as the source-probe spacing $x$ increases, to finally vanish at $x$=20 mm. On the contrary, the offset to the planned dwell positions remains at the same level over the entire displacement range of the stepping source (i.e., 10 cm), regardless of the source-probe spacing $x$ (cf. Fig \ref{fig:dpa}(a)).

%%%%%%%%%%%%%%%%%%%%%%%%%%%%%%%%%%%%%

\subsection{Dwell time verification}

\begin{figure}
\includegraphics[width=0.55\textwidth]{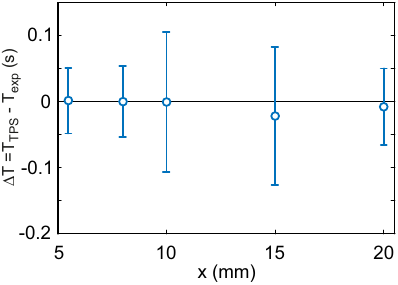}
\caption{Mismatch $\Delta T$ between the measured and planned dwell times ($ T_{exp}$ and $T_{TPS}$, respectively) as a function of the source-probe spacing $x$.}
\label{fig:dt}
\end{figure}

In Fig. \ref{fig:dt}, we report the analysis of the offset $\Delta T$ between the experimentally determined and expected dwell times ($T_{exp}$ and $T_{TPS}$, respectively) versus the source probe inter-catheter spacing $x$. The error bars show the mean and SD of the mismatch to the planned dwell times over the 40 dwell positions for each source-probe inter-catheter spacing $x$. The average of $\Delta T$ is calculated to be -0.006 $\pm$ 0.009 s. 

\section{Discussion}

\subsection{Detector characterization}

Our MPD shows an acceptable SNR for an accurate dwell time and position verification. At a source probe spacing $x$ of 20 mm, Therriault et al reported a SNR ranging from 13 to 23 with their multiprobe detector system involving three Photomultiplier tubes (PMT) as photometers \cite{therriault:mp13}. %The detection rate was not mentioned in this study. 
At the same source-probe spacing, we here measure  a SNR spanning from 33 to 39 with an sCMOS camera, a detection rate of 0.06 s and detection volumes that are 250 to 650 times smaller than those of Therriault's multiprobe. SNR could even be enhanced either by slightly broadening the scintillation cell and the fiber core (cf. Ref. \cite{gonod:pmb22}) or by replacing the sCMOS camera by seven PMTs which provide higher light detection sensitivity and an acceptable detection speed. However, seven PMTs may be more expensive than an sCMOS camera. Moreover, the performance of these ultrasensitive photometers can be seriously reduced after misuse. 

Beyond the opportunity of inserting a seven-probe detector within a brachytherapy needle, the advantage of the miniaturization of a water nonequivalent fiber dosimeter resides in a smaller volume averaging effect in the strong dose gradients of the radioactive source, as well as a minimum electron fluence perturbation.  

\subsection{Dwell position verification}

% Two-probe detector 

%Used in a two-probe configuration, our detection system shows optimum performances at interprobe spacings $\delta$z spanning from 15 mm and 35 mm. At this $\delta$z range, we observe the minimum values of the deviation $\Delta \bar{z}$ (Fig. \ref{fig:dpi}(a))  as well as the best trade-off between the fluctuations of the instant source positions $z_{exp}$ measured at various $x$ values (Fig.\ref{fig:dpi}(b)). 

\subsubsection{Optimal probe-to-probe spacing}

At an interprobe spacing $\delta z$ of 36.8 mm and at a source-probe spacing $x$=5.5 mm, our MPD used as a two-probe detector ensures a deviation to the planned dwell positions of 0.03 $\pm$ 0.15 mm. We here followed the measurement process depicted in Fig. \ref{fig:dpi}(b), with source positions expanding over 51.8 mm along (0z) (i.e., 36.8$\pm$7.5 mm). As a comparison, the deviation to the planned dwell positions is estimated to be 0.45 $\pm$ 0.3 mm (1SD) with the three-probe fiber detector of Linares et al \cite{linares:mp20}, at a source-probe spacing $x=5$ mm and a source activity of 10.73 Ci. Linares's detection system consists of three 17-mm spaced scintillation cells integrated at the end of an individual fiber detection line. Two reasons may explain the higher source-tracking accuracy of our two probe device %over Linares's three-probe detector
, despite detections of equivalent SNR. First, our probes show detection volumes that are two orders of magnitude smaller, thereby limiting the averaging effect in the steep dose gradients near an HDR-BT source. Second, spectral cross-talk between the three detection channels of Linares's set-up may limit measurement accuracy. In our case, the two detection channels are totally decoupled. 

With a 21-mm spaced four-probe detection system, Guiral et al demonstrated a mismatch to the planned dwell positions of 0.11$\pm$ 0.7 mm at a 0.1 s detection rate and along a 60 mm portion of source catheter. At a similar source probe spacing estimated to be of 20mm, we find a deviation to the planned dwell positions of 0.03$\pm$0.14 mm over dwell positions expanding over 50 mm. %With only two scintillation cells, we obtain a detection accuracy surpassing that of Guiral's set-up. 
Our detection volume being shrinked by 45 times regarding Guiral's, volume averaging effect is reduced, thereby improving dwell position verification.   

\subsubsection{Optimal algorithm for source tracking: two-probe versus all-probe detection strategies}

%compa two probes /all probes
Cascading scintillating probes along the source catheter allows to extend optimum source tracking capabilities over longer source paths. %With the above-studied two-probe detector (35-mm interprobe distance), the mismatch to the planned dwell position over the 10-cm travel range of the stepping source is of -0.01 $\pm$ 0.14 mm and 0.008 $\pm$ 0.14 mm at $x$ equal to 5.5 and 20 mm, respectively. 
%Although the above-discussed two-probe architecture shows good results, 
The source tracking accuracy reported above can be improved when the probe pair is dynamically chosen among the seven available probes of the detector to follow the source during the treatment delivery (all seven probes are sequentially involved in the treatment monitoring). The detection accuracy for the "two best gradient" surpass that of the  "two best intensity" method. At a 10 s dwell time, source tracking accuracy from these two methods is only slightly better than for the fixed two-probe system  but the gap should be noticeably enhanced for shorter dwell times, especially the sub-second regime. The two investigated methods could also show improved performances if a constant interprobe spacing was considered. %Probably a three-probe architecture would be sufficient to accurately monitor with these two methods a source stepping along a 10-cm-long catheter portion positionned at $x$ values up to four centimeter, or beyond.  

Source tracking accuracy is noticeably enhanced when the source position monitoring is realized via the accumulation of all the seven readout signals of the detection system. In that case, the deviation to the planned dwell positions is reduced to 0.029$\pm$0.078 mm and 0.02$\pm$0.19 mm at the source-probe spacings $x$ of 5.5 and 20 mm, respectively. At $x=5.5$ mm, we observe a maximum offset to the treatment plan of 0.12 mm. As a comparison, Linares et al reported a maximum deviation of 1.8 mm at $x=5$ mm and at similar source displacement range (10 cm) and dwell time (10s) \cite{linares:mp20}. At $x=20$ mm, Guiral et al found a discrepancy to the treatment plan of 0.11$\pm$ 0.7 mm with their four-probe detection system (at a dwell time of 5s) \cite{guiral:mp16}. The accuracy of the MPD in dwell position verification is fully compatible with the requirements of HDR Brachytherapy in terms of medical treatment and quality assurance \cite{fonseca:piro20,rivard:mp04,nath:mp97,kubo:mp98}.

%Analyse All-probe detection

Over the entire displacement range of the stepping source (i.e., 10 cm) and at a dwell time of 10 s, the "all-probe" detection approach provides a dwell position verification that is not affected by the fluctuations of the instant source position $\bar{z}_{exp}$, whatever the source-probe spacing $x$. Although the SD of $\bar{z}_{exp}$ varies by one order of magnitude during the treatment at $x=5.5$ mm (Fig. \ref{fig:dpa}(b)), the dwell position measurement remains at the same accuracy level (Fig. \ref{fig:dpa}(a)). The offset to the planned dwell position shows a narrow distribution of 0.002 $\pm$ 0.049 mm and a maximum value which does not exceed 0.12 mm. Such performances exceed those of competing multiprobe scintillator detectors \cite{linares:mp20}. %Over a similar range of source positions (10 cm), Linares's multiprobe detection system cannot ensure this accuracy steadiness \cite{linares:mp20}.
%At an identical dwell time of 10 s and at $x=5$ mm, Linares et al found a deviation to the planned dwell positions varying from 0 to 1.8 mm during a single treatment delivery \cite{linares:mp20}. %The range of measurement error is here enhanced by 15 times with regard to our multiprobe detection system. 
Note that all these results are obtained at a noticeably long dwell time. Measurements within a range of shorter dwell times (typically 0.1-10 s) will be studied in the future on some treatment plans used for instance in prostate brachytherapy. Preliminary results obtained with a single MSD are promising \cite{gonod:pmb22}.  
  
\subsection{Dwell time verification}

Across the source-probe spacings $x$ ranging from 5.5 to 20 mm, the total average of the differences $\Delta T$ between the measured and planned dwell times is of -0.006$\pm$0.009 s, at a detection rate of 0.06 s. At $x=20$ mm, $\Delta T$ is found to be -0.008 $\pm$ 0.058 s. With their four-probe detector, Guiral et al measured a difference $\Delta T$ of 0.05 $\pm$ 0.9 s at approximately similar source-probe spacing, a detection rate of 0.1 s and for 5-s dwell times \cite{guiral:mp16}. In their study, the source positions expanded over 6 cm within the source catheter, rather 10 cm as in our case. At $x=5.5$ mm, we find an offset $\Delta T$ of 0.0009 $\pm$ 0.0497 s.  As a comparison, Linares et al reported an offset to the planned dwell time of 0.33 $\pm$0.37 s at $x=5$ mm and a dwell times of 1s.  

It is noteworthy that the measurement errors of the dwell time  discussed here are all obtained from an edge detection in a staircase detection signal. In all the proposed methods, dwell times are defined as the elapsed time in between two successive signal edges, which are identified from a signal derivative calculation. Since all the considered dwell times in the above-cited studies are at least 50 fold longer than the integration time of the photometers used, one can assert that the measurement accuracy will not significantly change as the dwell time increases. We recently verified this property in a single probe detection \cite{gonod:pmb22}. Therefore, the results from Guiral's and Linares's multiprobe detectors obtained with 5 s and 1 s dwell times, respectively, can be directly compared to our results measured at a dwell time of 10 s. As an example, the lower measurement accuracy of Linares's approach may be partly explained by their shorter 1 mm inter-dwell spacing, leading to noticeably smaller temporal edges (given the strong dose gradients involved), rather than the use of a shorter dwell time of 1s.

\subsection{Clinical use}

Our multiprobe detector is compatible with clinical applications. It indeed consists of biocompatible elements and it is sufficiently narrow to be inserted in a BT needle or in a catheter. As a preliminary step, we successfully positioned our detector in a one-millimeter wide sealed encapsulation pipe made of PEEK material. 

\textit{In vivo} applications forbid the use of our motorized stage for probe positioning.  In gynecologic BT, the probe and the source would be inserted in two parallel catheters of an applicator. The inter-catheter spacing would be precisely known, as in the case of the present study. In prostate BT, the probe and the source are inserted in two independent needles which are implanted manually in a patient. Therefore, $x$ and $z$ coordinates (cf. Fig. \ref{fig:exp}(d)) are usually coupled since the needles are rarely implanted perfectly parallel from each other, due to operational uncertainties. $x$ and $z$ coordinates of the source relatively to the probe can however be simultaneously determined by various triangulation approaches rendered possible  by our seven-probe detection system (see for instance Ref. \cite{linares:mp20}). Source tracking via a triangulation process requires a refine 2D calibration plot of the system, which does not represent a challenge here \cite{gonod:pmb22}. Moreover, parallel IVD from a pair of multi-probes detectors connected to the same camera and inserted in two different needles would allow a 3D positioning of the source by triangulation \cite{linares:mp21} with unprecedented accuracy and minimum equipment. 

%\begin{figure}
%\begin{center}
%\includegraphics[width=0.6\textwidth]{triangulation_8mm.pdf}
%\caption{}\label{fig:triang}
%\end{center}
%\end{figure}

%As a first demonstration of a simultaneous determination of the $x$ and $z$ source coordinate by triangulation is shown in Fig. \ref{fig:triang}. From a local refine 2D calibration of the detector over $x$ ranging from 5.5 mm to  10 mm and $z$ spanning over 10 cm, we realized 2D source tracking for a planned source-probe spacing of 8 mm. We repeat the dwell position verification from an "all-probe" detection but the 2D calibration plot as a position reference, instead of the the 1D plot of Fig. \ref{fig:calib}. We obtain a deviation to the planned source coordinates $x$ and $z$ of ?? $\pm$ ?? mm and ?? $\pm$ ?? mm, respectively, which far surpasses preceding source tracking performances based on triangulation \cite{linares:mp20}. 

\section{Conclusion}

We have demonstrated in a water phantom a monitoring device for HDR-BT based on a seven-probe scintillator dosimeter coupled to an sCMOS camera. Being engineered at the end of a narrow 270 $\mu$m diameter fiber bundle, our miniaturized probes combine high spatial resolution and high detection speed while ensuring a minimum perturbation of the therapeuty, even if water nonequivalent (inorganic) materials are used. Moreover, the overall dosimeter is totally free from inter-probe cross-talk. The use of an sCMOS camera, rather than seven photomultiplier tubes of higher sensitivity offers the possibility of a simultaneous parallel readout of the seven probe signals in a simple and low cost architecture that is well adapted to a clinical use.  The smaller SNR of the camera is compensated by a higher probe detection efficiency enabled by our concept of an IVD micro-pixel based on a nano-optical interface in between scintillators and a fiber \cite{suarez:ox19}. 

First, we found a range of probe-to-probe spacings which minimizes source tracking uncertainties. This will be an important information for future MPD designs. We then studied and compared three different source tracking algorithms among the large panel of possibilities offered by our seven-probe system. %In a time resolved dosimetry from a single pair of 36.8-mm spaced probes, the offset to the planned dwell position along the source catheter is of 0.008 $\pm$ 0.14 mm at a source-probe spacing $x$ of 20 mm and over a source displacement range of 10 cm (along (0z)). Accuracy is a little improved when the probe pairs are dynamically chosen during the treatment delivery to follow the stepping source. 
The best detection approach was found by adding the parallel readout signals from all the probes of the detector. Realizing a source tracking based on this overall accumulated readout signal led to an offset to the planned dwell position as small as 0.01 $\pm$ 0.14 mm and 0.02 $\pm$ 0.29 mm over a 10-cm long source displacement in the source catheter and at spacings between source and probe catheters of 5.5 and 40 mm, respectively. Using this method, we also measured deviations to the planned dwell time of -0.006$\pm$0.009 s and  -0.008 $\pm$ 0.058 s, at source-probe spacings $x$ of 5.5 mm and 20 mm, respectively (detection rate of 0.6 s). All the studied configurations were found to surpass current fiber-integrated multiprobes detection systems. The next steps will be to test our detection system with various treatment plans used for instance in prostate brachytherapy. The detection performances demonstrated here need to be assessed at shorter dwell times down to a fraction of a second. Triangulation approach will also be realized to simultaneously define the dwell time and the 2D coordinates $x$ and $z$ of a stepping HDR-BT source with unprecedented accuracy. 

\section*{Acknowledgments}
The authors thank Lionel Pazart, Thomas Lihoreau and Karine Charriere for helpful discussions. This study is funded by the SAYENS Agency, the French Agency of Research (contract ANR-18-CE42-0016),  the Region "Bourgogne Franche-Comte" and the EIPHI Graduate School (contract ANR-17-EURE-0002). This work is supported by the French RENATECH network and its FEMTO-ST technological facility.

%\bibliography{base_biblio}
%\bibliographystyle{medphy.bst}

\section*{References}
\addcontentsline{toc}{section}{\numberline{}References}

% Following assumes you are using bibtex. However, for submission to the
% journal you MUST explicitly INCLUDE THE REFERENCES IN THE TEX FILE. 
% In that case you need the following

\end{document}